\documentclass{ws-rv9x6}
\usepackage{siunitx}
\usepackage{subfigure}   
\usepackage{ws-rv-thm}   
\usepackage{ws-rv-van}   
\makeindex

\begin{document}

\chapter{Prospects and Results from the AFP Detector in ATLAS}

\author[G.P. Gach]{Grzegorz Gach\\On behalf of the ATLAS Collaboration}

\address{AGH University of Science and Technology\\
Faculty of Physics and Applied Computer Science\\
al. Mickiewicza 30, PL-30059 Cracow, Poland\\
grzegorz.gach@cern.ch}

\begin{abstract}
  In 2016 one arm of the AFP detector was installed and first data
  have been taken. In parallel with integration of the AFP subdetector
  into the ATLAS TDAQ and DCS systems, beam tests and preparations for
  the installation of the 2$^{\textrm{nd}}$ arm are performed. In this
  report, a status of the AFP project in the ATLAS experiment is
  discussed.
\end{abstract}

\body

\section{Introduction}\label{sec:intro}

ATLAS\cite{Aad:2008zzm} is a general purpose detector operating at the
LHC\cite{Evans:2008zzb}\,. Apart from the main apparatus a new group of
forward detectors is installed in order to extend the scope of
possible measurements. The existing Zero Degree
Calorimeter\cite{Aad:2008zzm} mounted \SI{140}{\metre} away from the
nominal interaction point (IP) is dedicated to the measurement of the
forward neutrons and photons. The Absolute Luminosity For ATLAS
(ALFA)\cite{AbdelKhalek:2016tiv} detector, situated about
\SI{240}{\metre} from IP, is devoted to the measurement of elastic
scattering. The latest detector is named ATLAS Forward Proton
(AFP)\cite{Adamczyk:2017378}\,. It is dedicated to studies of
diffractive processes and its main purpose is to measure forward
protons.

\section{The AFP Detector}
\label{sec:afp}

The AFP detector is designed to consist of four stations, two on each
side of ATLAS along the beam-pipe. The stations are located
\SI{\pm{}206}{\metre} and \SI{\pm{} 214}{\metre} away from the ATLAS
IP. Each station is equipped with Roman Pots which allow the
horizontal movement of the detector components and their insertion
into the beam-pipe. This enables precise positioning of the detector
in the vicinity of the beam. Positions of the stations and detectors
result in acceptance of protons with relative energy loss of about
${\xi = (E - E')/E \in (0.02, 0.12)}$, where $E$ and $E'$ are incoming
and outgoing proton energies, respectively.

Each station is designed to house four planes of the tracking
detectors. Far stations (i.e. the ones placed at \SI{\pm{} 214}{\metre})
will be equipped with time-of-flight counters.

Silicon pixel detectors are used for tracking. Each plane contains
$336\times80$ pixels of size
$50\times250$\si{\micro\metre\squared}. The modules are similar to the
ones used in ATLAS IBL\cite{Capeans:1291633} and are radiation
hard. The detectors are tilted by \SI{14}{\degree} with respect to the
vertical direction, which makes that majority of the protons pass two
pixels.

In the Technical Design Report\cite{Adamczyk:2017378} the
time-of-flight counters are stated to have time resolution of
\SI{10}{\pico\second} or better and efficiency of at least
\SI{90}{\percent}. This enables measurements in the environment of
about 50 interactions per bunch crossing on average and allows to
identify the vertex from which the proton measured in AFP originates
with resolution of \SI{2.1}{\milli\metre} in the longitudinal
direction. Moreover, the detectors have to be fast enough to provide
the trigger signal, which is challenging due to their location at
\SI{214}{\metre} away from the ATLAS detector. The main purpose of the
time-of-flight detectors is to identify the vertex from which the
proton measured in AFP originates. Thus they are crucial in runs with
many interactions per bunch crossing (high pile-up) in standard runs,
in which AFP is expected to take part in 2017.

AFP installation is staged. The first stage during which one arm of
the AFP detector ``AFP~0+2'' was installed took place before data
taking in 2016. The installation of the other arm, ``AFP~2+2'', is
scheduled for the beginning of 2017.

\section{AFP 0+2}
\label{sec:afpZeroTwo}

The first stage of the AFP detector installation covered placing two
stations in the LHC tunnel. Both stations were situated on the same
side of ATLAS. They were equipped only with tracking detectors. The
stations had to pass the LHC qualification, because the detectors are
inserted into the beam-pipe. Additionally, the detector control (DCS),
data acquisition (DAQ) as well as trigger systems had to be integrated
with the ATLAS central systems. As a result of successful installation
and almost flawless detector operation it was possible to collect
first data for physics analysis already in 2016, although it was not
initially planned.

\begin{figure}
\centerline{\includegraphics[width=0.67\textwidth]{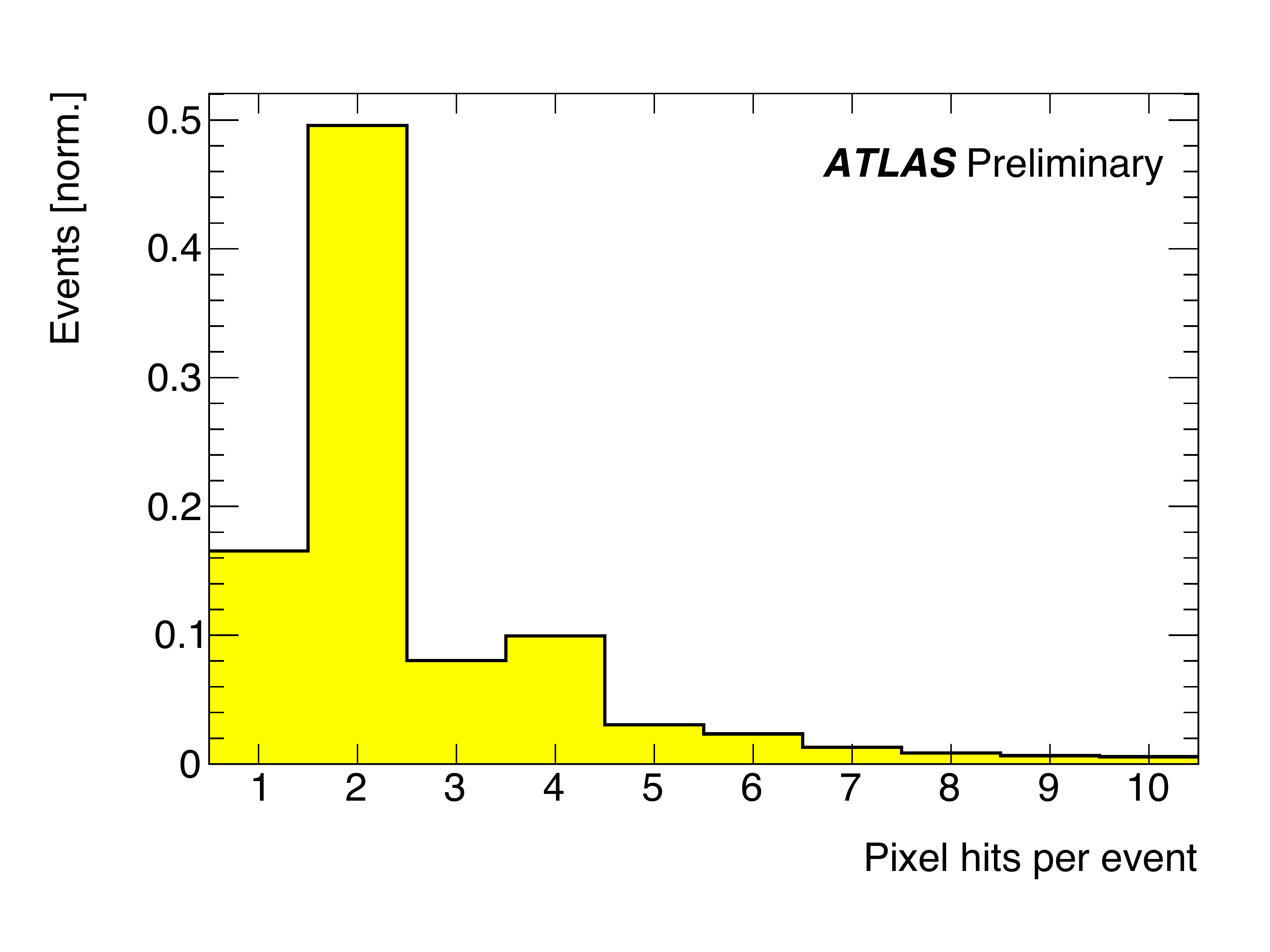}}
\caption{Number of hits in a plane based on 2016 data.\cite{fwdURL}
}\label{fig:afpPlanesHits}
\end{figure}

\begin{figure}
\centerline{\includegraphics[width=0.75\textwidth]{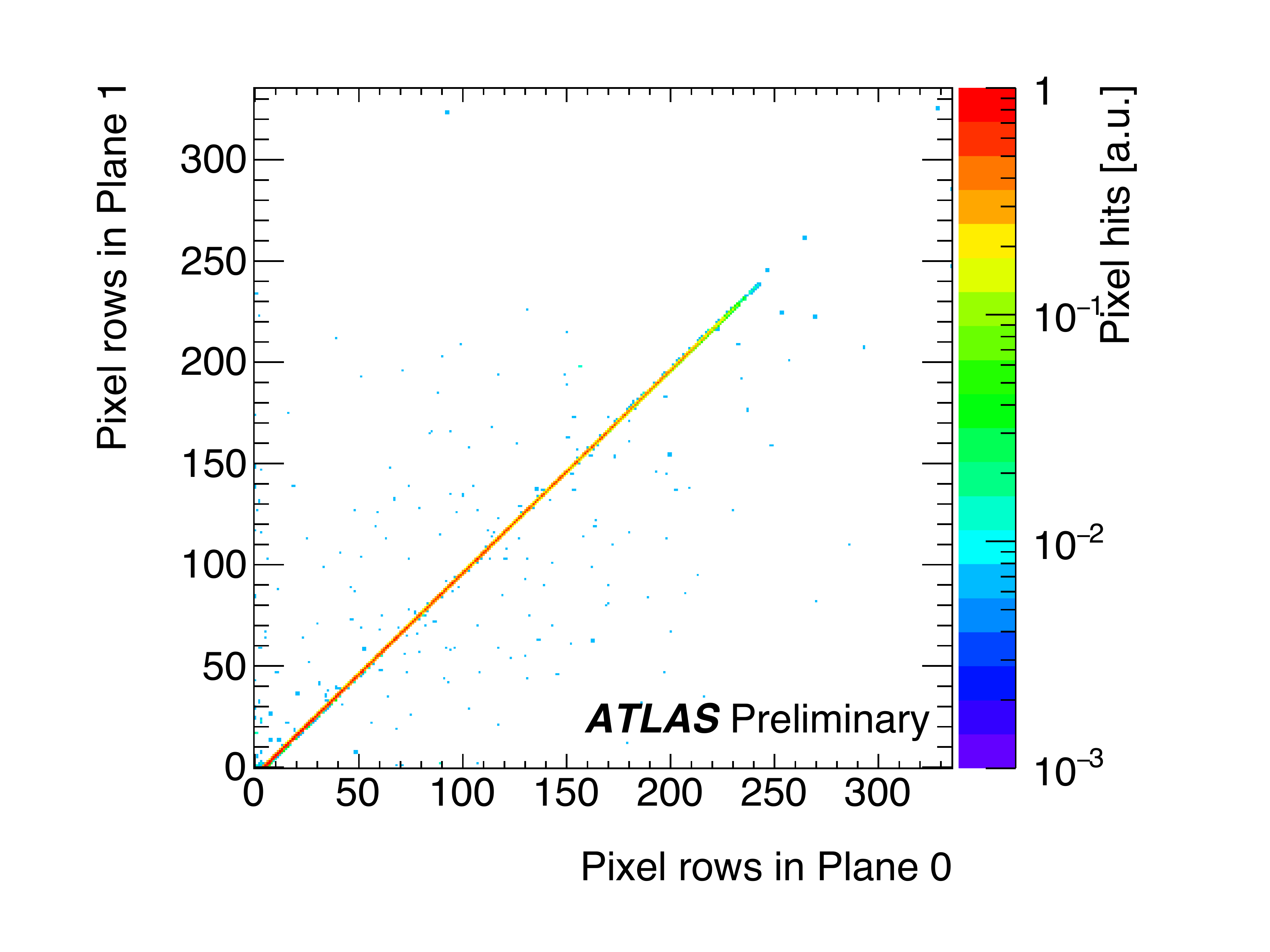}}
\caption{Correlation of pixel rows between first and second planes of
  tracking detectors based on 2016 data.\cite{fwdURL}
}\label{fig:afpPlanesCorr}
\end{figure}

In 2016 AFP participated in two physics runs. In these special runs
the pile-up was low. In total almost \SI{500}{\per\nano\barn} of
integrated luminosity was collected. The detectors performed very
well. In Fig.~\ref{fig:afpPlanesHits} the number of hits per silicon
detector plane based on the 2016 data is presented. In most of the
events the signal is registered in two pixels which is exactly what
was expected due to the tilt of the silicon planes. A strong
correlation between the neighbouring planes can be observed in
Fig.~\ref{fig:afpPlanesCorr} where the pixel row index in the second
plane versus the first plane is plotted for events with up to two
pixel hits in a layer.

A natural physics program for this configuration contains the studies of
single diffractive dissociation (SD). In this process a forward proton
and a hadronic system well separated in rapidity are present in the
final state. The cross section of this process is relatively large
which enables studies using the low pile-up data. So far the ATLAS
collaboration did not publish any result of the SD measurement with
identification of a forward proton, which allows to distinguish
between single diffractive and low-mass double diffractive
dissociations. The AFP detector delivers a unique possibility of
performing such measurements in the kinematic region of the relative
proton energy loss in the range $\xi \in (0.02, 012)$. Moreover, it
will provide access to so far unavailable kinematic variables like the
four-momentum transfer in the proton vertex $t = (P - P')^2$, where
$P$ and $P'$ are the incoming and outgoing proton four-momenta,
respectively.


\section{AFP 2+2}
\label{sec:afpTwoTwo}

The final AFP installation is scheduled for early 2017. It includes
placing of two new stations in the tunnel, installation or replacement
of the current tracking and time-of-flight detectors. These actions
have to be followed by the LHC qualification procedure. Additionally,
the time-of-flight detectors have to be integrated with the ATLAS
trigger system. The full instrumentation of AFP will allow data taking
in special low pile-up runs as well as in standard runs with high
pile-up.

This configuration, just like the AFP~0+2 one, is good for studies of
single diffractive dissociation in special runs with clean
environment. However, the possibility to measure forward protons on
both sides and collect high integrated luminosity in standard runs
gives access to central diffractive processes with small
cross-sections.  In addition, information about event's full
kinematics is provided. A reach diffractive physics program, including
exclusive dijet production or central diffractive photon-jet
production, aims at better understanding of the nature and structure
of the colour singlet exchange. However, the program is not limited to
this kind of studies only. There are also plans to investigate
photon-induced processes for which the proton tag provides a very good
background rejection. The exclusive lepton production
measurements\cite{Aad:2015bwa} can be improved. The
possibility to discover new physics lies e.g. in the studies of
anomalous quartic couplings.

\section{Summary}
\label{sec:summary}

All operations of the AFP detector, dedicated to measurements of the
forward protons mainly originating from diffractive interactions, are
on schedule. The first stage, AFP 0+2, was finished in 2016 and the
data for physics analysis were collected in special runs with low
pile-up. The second and final installation stage, AFP 2+2, will be
finished in 2017. The timing detectors installed then will allow the
measurements of the forward protons in standard runs with high
pile-up, thus providing access to central diffractive processes
characterised by small cross-sections. The usefulness of the AFP
detector is reflected in its reach physics program presented in the
``Technical Design Report for the ATLAS Forward Proton
Detector''\cite{Adamczyk:2017378}\,.

\section{Acknowledgement}

This work was partly supported by the National Science Centre of
Poland under contract No. UMO-135 2015/18/M/ST2/00098 and by PL-Grid
Infrastructure.

\bibliographystyle{ws-rv-van}
\bibliography{AFP}


\end{document}